# Turbulence compressibility reduction with helicity

Yan Yang (杨焱)[1] and Jian-Zhou Zhu (朱建州)[2, a)]

[1)] *LHD, Institute of Mechanics, Chinese Academy of Sciences, Beijing 100190, China*

[2)] *Su-Cheng Centre for Fundamental and Interdisciplinary Sciences, Gaochun, Nanjing 211316, China, and, Fluid Institute, Nanchang University, China.*

Numerical test of isotropic turbulence compressibility reduction with helicity in a cyclic box is performed. The ratios of compressibility-relevant-mode spectra over those of kinetic energy present power laws at large wavenumbers in the dissipation range, indicating a common difference of 11/15 in the exponents of the algebraic prefactor of the nonhelical power spectra over those of helical ones. Our results being not derived from the shapes of the spectra themselves, the implied information about the helicity effect on the complex singularities of the discretized dynamical system can still be of reasonable value for insights of the Navier-Stokes equation, although the high-order finite difference scheme used for computation may not be as accurate in dissipation range as the state-of-the-art of incompressible turbulence with pseudo-spectral method. Possible applications in controlling flows, for the purposes of, say, decreasing turbulence noise, are also discussed according to the spectral fluctuations.

[a)] Electronic mail: jz@sccfis.org

## I. INTRODUCTION

The 'helicity effect' in classical fluids and plasmas[1–3] has long been studied, but few clear messages are available for practical application, say, in the problems of aeroacoustics of wind engineering or aeronautics, to the best of our knowledge. A recent effort has been made with statistical analysis,[4] updating that of Kraichnan[5] (K55) with the information of helicity, which indicates the possibility that, compared to nonhelical flows, a turbulence with (more) helicity would bear less proportions of the energy of compressibility relevant fluctuations, such as those of the compressive, density and internal-energy modes. Such flow compressibility reduction has been termed the 'fastening' effect of helicity, and appears, most directly, of potential applications in turbulence noise and heating problems.

Further studies have been performed, as done[6] with the association via boost to a rotating frame to the mechanics or geometry in the compressible Taylor-Proudman theorem[7] for a particular type of helical field which is locally generic and finding some universality, say, in plasma flows. Preliminary test in some low-Reynolds-number, thus essentially laminar, flows also appear to be consistent with the above mentioned 'fastening' notion.[8]

In this communication we will present the first test of the fastening notion in fully-developed turbulence with reasonably high Reynolds number, followed by further analyses with the purpose of deeper mathematical understanding and practical applications: such fundamental research, just to mention the aeroacoustic aspect,[9] is indeed relevant in civil engineering, aeronautics, rapid public transportation and car transportation.[10,11]

An important application would be in the design problem (e.g., shape of airfoil or wing[12] and aeroacoustic and aerodynamic optimization of propeller blades,[13] among other objectives such as those combined with trajectory-based operations for enhancing the predictability of air traffic and flight efficiency with the idea of helicity controll to reduce flow noise. Without substantial experiments in laboratory or in silico, such discussions would be of speculative nature but are based on solid results.

Thus, the key results are on fundamental turbulence, but our purposes also include applications, especially the optimal design of, say, aircraft with the objective of reducing (turbulence) noise, the latter being of course based on the former. We first lay out the formulation of the problem and the methods of analysis, together with a simple variational argument, in Sec. II; the comparisons of the time-averaged spectra with and without helic-ity injection are then offered in Sec. III followed by Sec. IV for further measurements and discussions associated to the dissipation-range behavior and complex-singularity properties, and then by Sec. V with insights from the additional information of fluctuations around the time averages for our expectations of future relevant studies.

## II. BASICS AND METHODS

Let $\boldsymbol{u}$, $\rho$, $\eta$ and $c$ represent the velocity, density, viscosity and velocity of sound in a compressible flow, and the pressure $p$ given by the adiabatic (actually *isothermal*) relation

$$p = c^2\rho \text{ and } \rho = \rho_0 e^{\zeta} \text{ with an equilibrium } \rho_0. \tag{1}$$

Both $c$ and $\rho_0$ may be set to be unit ($= 1$), but sometimes we still let them present explicitly to emphasize the physical reality. The usage of the logarithmic variable[5] $\zeta$ is convenient for the quadratic expressions of the approximate energy for small $\zeta$ and is also exploited in many numerical solvers[8,14,15] to naturally preserve the positiveness of $\rho$ and obey the governing equation of self-consistent dimensionally reduced flows.[16,17] The equations of motion then write

$$\partial_t\zeta + \zeta_{,\alpha}u_\alpha + u_{\alpha,\alpha} = 0, \tag{2}$$

$$\partial_t u_\lambda + u_\sigma u_{\lambda,\sigma} + c^2\zeta_{,\lambda} - \eta\theta_{\lambda\sigma,\sigma} = 0, \tag{3}$$

where $\theta_{\alpha\beta} = u_{\alpha,\beta} + u_{\beta,\alpha} - \frac{2}{3}\delta^\alpha_\beta u_{\sigma,\sigma}$, $(\bullet)_{,\gamma} = \partial(\bullet)/\partial x^\gamma$. Working in a cyclic box of dimension $2\pi$ with $V = [0, 2\pi)^3$ and applying Fourier representation for all the dynamical variables $v(\boldsymbol{r}) \to \hat{v}(\boldsymbol{k})$, say, $\boldsymbol{u}(\boldsymbol{r}) = \sum_{\boldsymbol{k}} \hat{\boldsymbol{u}}(\boldsymbol{k}) \exp\{\hat{i}\boldsymbol{k}\cdot\boldsymbol{r}\}$ with $\hat{i}^2 = -1$, we can construct a phase space by the real and imaginary parts of $v$s and Galerkin truncation, imposing all modes with $k = |\boldsymbol{k}|$ greater than some cut-off value $K$ to be zero, may be performed, which does not change the approximated conservation of energy due to its being quadratic and diagonal in $\boldsymbol{k}$.[18] We can show that the flow in such a phase space is incompressible (the Liouville theorem) in the inviscid case. Reasoning with the $H$-theorem, K55 expects an ensemble of systems tend towards an absolute statistical equilibrium state.

The helicity, or mean helicity density in the volume $\mathcal{V}$,

$$\mathcal{H} := \int \frac{\boldsymbol{\omega}\cdot\boldsymbol{u}d^3\boldsymbol{r}}{2\mathcal{V}}, \text{ with the vorticity } \boldsymbol{\omega} = \nabla\times\boldsymbol{u}, \tag{4}$$

is an ideal barotropic invariant[2]. Using the helical representation[19]

$$\hat{\boldsymbol{u}}(\boldsymbol{k}) = \hat{u}_+(\boldsymbol{k})\hat{\boldsymbol{h}}_+(\boldsymbol{k}) + \hat{u}_-(\boldsymbol{k})\hat{\boldsymbol{h}}_-(\boldsymbol{k}) + \hat{u}_|(\boldsymbol{k})\boldsymbol{k}/k, \tag{5}$$

with the property of the helical basis $\hat{i}\boldsymbol{k} \times \hat{\boldsymbol{h}}_s = sk\hat{\boldsymbol{h}}_s$ (for $\hat{i}^2 = -1$ and $s = \pm$), and that,

$$\mathcal{E} = \frac{1}{2}\sum_{\boldsymbol{k}} |\hat{u}_+|^2 + |\hat{u}_-|^2 + |\hat{u}_||^2 + c^2|\hat{\zeta}|^2, \tag{6}$$

$$\mathcal{H} = \frac{1}{2}\sum_{\boldsymbol{k}} k|\hat{u}_+|^2 - k|\hat{u}_-|^2, \tag{7}$$

we then can obtain with the canonical distribution $\sim \exp\{-(\alpha\mathcal{E} + \beta\mathcal{H})\}$ the absolute-equilibrium modal spectra which indicate that the proportion of compressibility-relevant-mode energy may be reduced with helicity.[4]

A probably even simpler variational argument supplemented in Ref. 17 is the following. Suppose there be a maximally helical flow with given energy. With the lagrangian multiplier $r$ to form $\mathcal{H} + r\mathcal{E}$, the constrained extremization (variations with respect to, respectively, $u_+$, $u_-$, $u_|$ and $\zeta$) requires that

$$\hat{u}_| = \hat{\zeta} = 0 = \hat{u}_- \text{ and } r = -k \text{ for } \hat{u}_+(\boldsymbol{k}) \neq 0, \tag{8}$$

or that

$$\hat{u}_| = \hat{\zeta} = 0 = \hat{u}_+ \text{ and } r = k \text{ for } \hat{u}_-(\boldsymbol{k}) \neq 0. \tag{9}$$

Thus, it is expected that in general flows larger $|\mathcal{H}|$ corresponds to less non-vortical-mode excitations, i.e., 'less compressible' or 'more strongly fastened'.

We performed direct numerical simulations of the compressible Navier-Stokes equations (NSE) in a cyclic box, using Pencil Code[15] with sixth-order compact scheme for spatial finite difference and third-order Runge-Kutta time marching. And, as summarized in Fig. 1 and as will be further explained below, we also introduce the one-dimensional (1D) spectra

$$E_- := \sum_{|\boldsymbol{k}|=k} \frac{|\hat{u}_-|^2}{2},\ E_+ := \sum_{|\boldsymbol{k}|=k} \frac{|\hat{u}_+|^2}{2},\ E_| := \sum_{|\boldsymbol{k}|=k} \frac{|\hat{u}_||^2}{2} \text{ and } E(k) := E_- + E_+ + E_|. \tag{10}$$

A specific statistical conjecture is that $E_|$ is of smaller fraction for a helical, compared to the nonhelical, system, as realized in Fig. 1 by, respectively,

$$E_+ \neq E_- \text{ [thus a nonvanishing helicity spectum } H(k) = kE_+ - kE_-] \text{ and } E_+ = E_-. \tag{11}$$

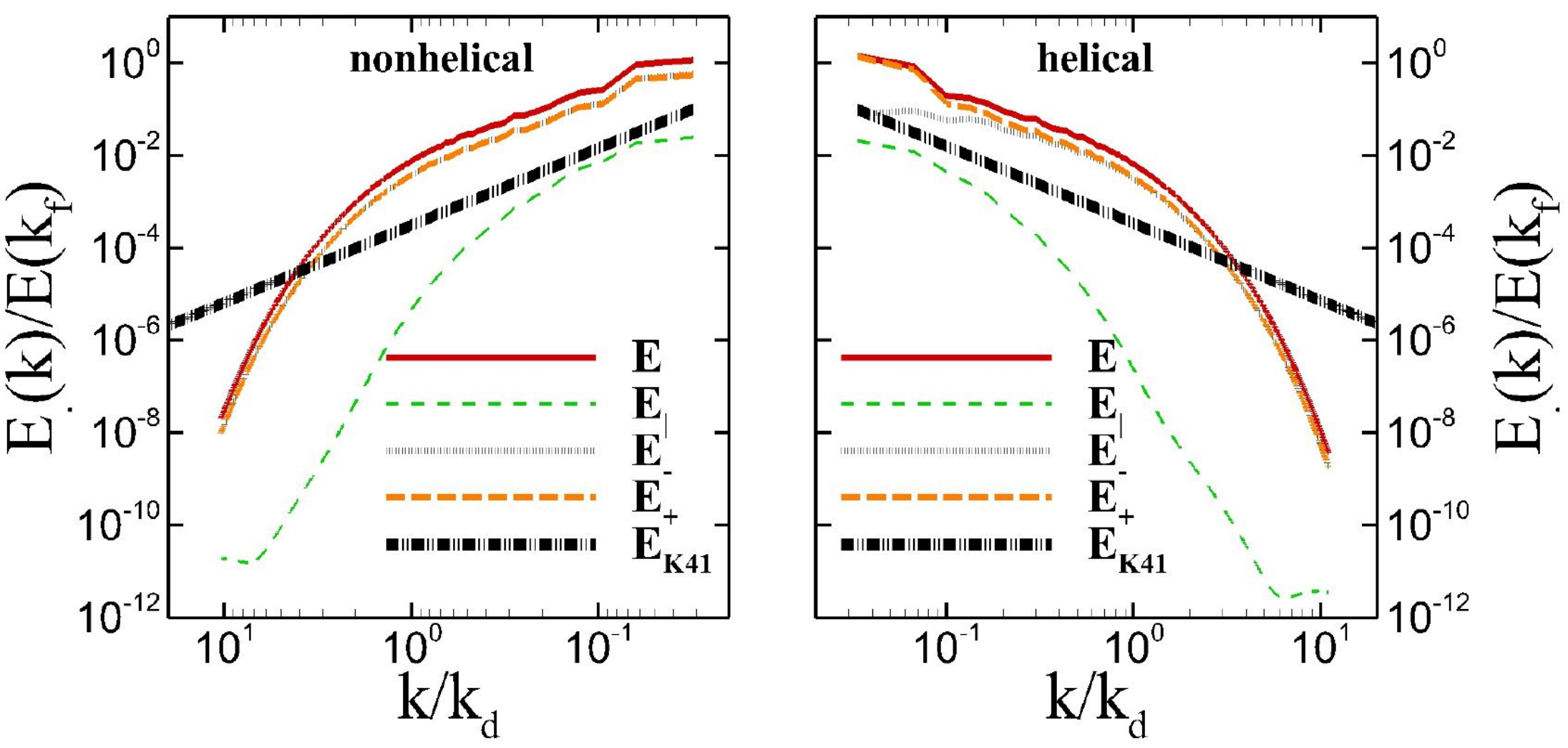

FIG. 1. The normalized power spectra (time-averaged) of kinematic energy and its components, for the cases with and without helicity injections: the nonhelical case shows basically identical spectra of the left- and right-handed vortical modes, while the helical case presents larger right-handed vortical-mode spectra at small $k$. Due to the numerical and computer noise problem, the results of compressive (and density: not shown) modes are not reliable at the high $k$ ends.

## III. RESULTS

Note also that even with the same viscosity coefficients and strengths of forcing (helical or not), since the helicity can change the dynamical properties,[18,20] energy levels and precise shapes of the spectra from two simulations with and without helicity injection are different. Thus, a naive comparison of helical and nonhelical turbulence without appropriate normalization would not be very illuminating. In our simulations, the isotropic and $\delta$-correlated stochastic accelerations were added at around wavenumber $k_f = 1.54$, and the solenoidal forcing scheme using essentially the helical representation as in Eq. (5) in Ref. 21 was applied for controlling the degree of chirality of the accelerations (and thus the respective flows), the latter being the genuine difference between the two runs. [Forcing only on the transversal modes is mainly from the consideration of simplicity and fairness in the comparison, though we believe other forcing schemes with (statistically) appropriate drivings on the longitudinal or even the other compressibility relevant modes, not exhausted in this note, should also work. We adopted for this particular test the setup of the magnetohydrodynamic one of Ref. 21 (thus a particularly detailed explanation of the method would be redundant), with the magnetic field removed.]

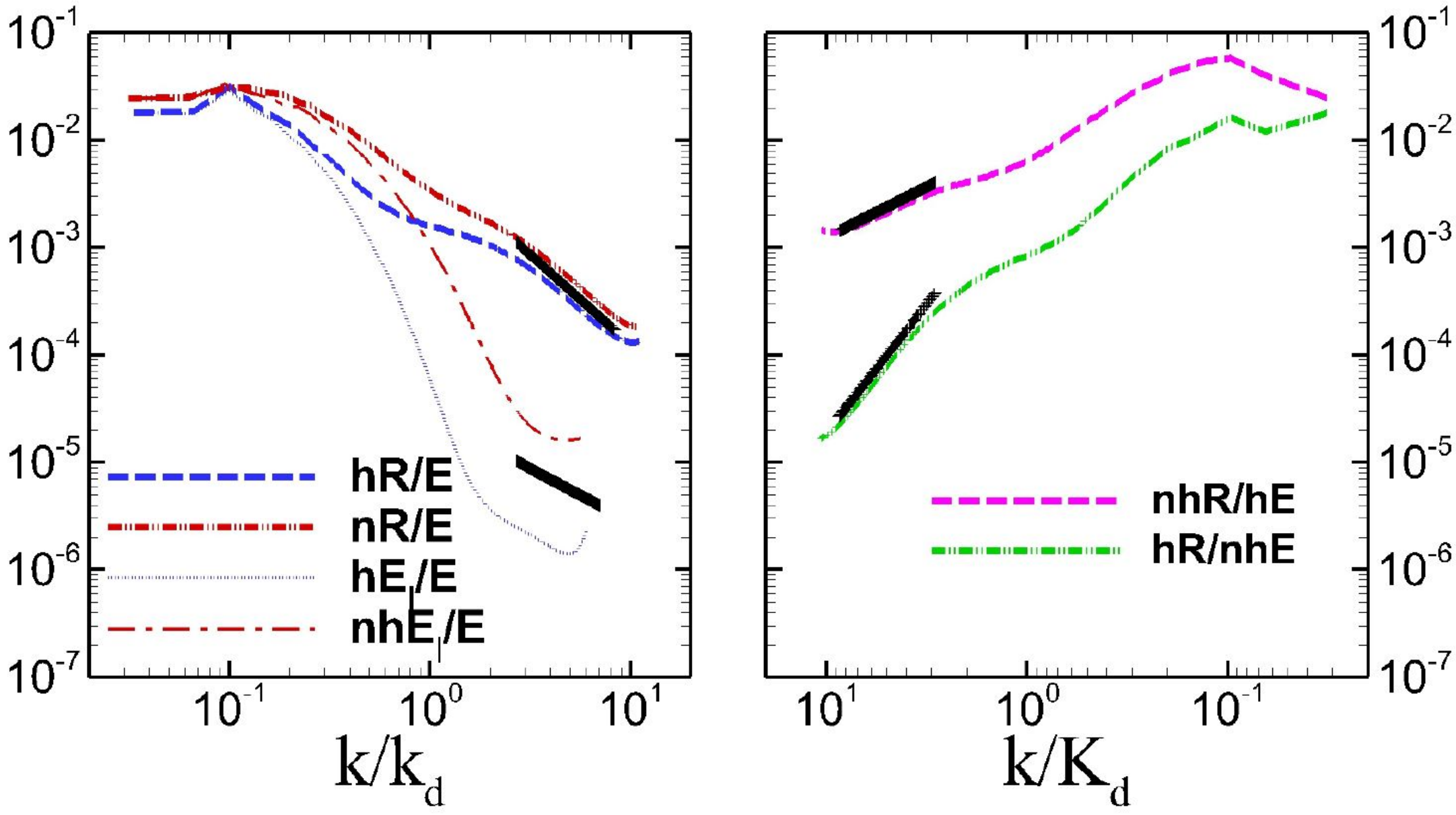

FIG. 2. The time-averaged compressible- and density-mode power spectra of the helical and nonhelical cases: the left panel is for the "self-normalization", i.e., $E_{|}(k)$ and $R(k)$ of respectively helical ('$hE_{|}/E$' and '$hR/E$') and nonhelical ('$nhE_{|}/E$' and '$nR/E$') cases normalized by $E(k)$ of the same case ($k/k_d$ of the same fashion); the right panel is for the "cross-normalization", i.e., helical $R$ normalized by nonhelical $E$ ('$hR/nhE$') and nonhelical $R$ normalized by helical $E$ ('$nhR/hE$'), while $k$ is normalized by the common $K_d$.

Both cases are of the root mean square turbulent Mach number $M_t \approx 0.4$ and of Reynolds numbers $Re \approx 250$ in terms of Taylor micro-scales [the '$_n$(onhelical)' case is slightly (6%) higher than the '$_h$(elical)' one, $Re_n = 257$ versus $Re_h = 243$; and, for dissipation wavenumbers, $k_{dn} = 31.7$ versus $k_{dh} = 30$]. Accordingly, two normalizations are then designed for physical analysis: Fig. 1 presents various spectra normalized by $E(k_f)$, including the (unnormalized) Kolmogorov spectrum ('K41')[22] $\propto k^{-5/3}$ for reference, with $k_f = 1.54$ for the reason that all modes of $|\boldsymbol{k}| = \sqrt{2}$ and $\sqrt{3}$ are forced; while the spectra of the density and compressive modes normalized by $E(k)$ are plotted in Fig. 2, left panel, both showing consistent results of reduced compressibility with helicity.

Relevant to (aero)acoustic noise measured by the pressure fluctuation $p'$, it is typical to gain a benefit

$$\Delta SPL_{rms} := 20 \log_{10} \frac{p'^{nonhelical}_{rms}}{p'^{helical}_{rms}} \approx 3.0 \text{ dB}, \tag{12}$$

evaluated here with the root mean squares of the pressure fluctuation of the helical and nonhelical cases computed from the normalized time-averaged power spectra in our simulations. The benefits fluctuating with time and depending on the scales, such a gain of 3.0 dB should however mainly be considered as a more definite 'proof of the concept' rather than a solid number for practical guidance, though it is not impossible that our fundamental results may also further be applied to various situations, from understanding to control, and, from nature to laboratory and to industry (c.f., e.g., Refs. 10 and 11 for relevant problems).

## IV. FURTHER DISCUSSIONS

The computation and measurement of the dissipation-range fluctuations are highly nontrivial, due to the smallness of both space-time scales and amplitudes of the fluctuations (especially those of the compressive and density modes), and the current high-order numerical schemes, 'almost spectral accuracy'[15,21] though, probably won't be able to produce a reliably accurate shapes of the spectra of turbulence at this regime [see also, e.g., Refs. 23 for the great efforts, but no clear success yet, to determine the dissipation range spectra of incompressible flow from (pseudo-)spectral simulations which are in principle of most numerical reliability with spectral accuracy]. However, the phenomena of the ordinary-difference dynamical system defined by the numerical scheme to discretize NSE by themselves are of course interesting; and, it is possible that differences between the computed spectra [here $R(k)$ and $E(k)$] and those of the original NSE be reasonably removed or greatly reduced in the ratios and that Fig. 2 makes sense for demonstrating the helicity effect on turbulence. Thus, in this and the next sections we slightly extend the further discussions in Ref. 17 for deeper insights.

We observe in Fig. 2, as designated by short straight lines in the dissipation ranges,

$$E_{|}(k)/E(k) \propto k^{-1} \tag{13}$$

$$\text{and } R(k)/E(k) \propto k^{-5/3}, \tag{14}$$

a new *universality* agreed by the helical and nonhelical cases. The above results indicate that, in either the helical or nonhelical case, the $E_{|}(k)$ and $R(k)$ are different from the $E(k)$ only up to power-law prefactors in the (stretched-)exponential decay.

We may then accordingly postulate the (asymptotic) *ansatz* for the four spectra, helical and nonhelical $R(k)$ and $E(k)$ (the $E_|$ relevant behaviors are consistent with but not as clean or as strongly indicative for such a postulation: see below)

$$\propto k^{\alpha} F(k) \ . \tag{15}$$

$F(k)$ is some empirical (stretched) exponential function, with the argument $k$ usually normalized by $k_d$ in the K41 universality phenomenology, or both further 'renormalized' with $\ln Re$ in the multifractal-universality phenomenology.[22–24]

The observed Eq. (13) in this dissipation range indicates that, with again '$_h$(elical)' and '$_n$(onhelical)' and the self-evident superscripts,

$$F_h^{E_|}(k) = F_h^E(k),\ F_n^{E_|}(k) = F_n^E(k), \tag{16}$$

$$\alpha_h^{E_|} - \alpha_h^E = -5/3 = \alpha_n^{E_|} - \alpha_n^E. \tag{17}$$

And Eq. (14) leads to

$$F_h^R(k) = F_h^E(k),\ F_n^R(k) = F_n^E(k), \tag{18}$$

$$\alpha_h^R - \alpha_h^E = -5/3 = \alpha_n^R - \alpha_n^E. \tag{19}$$

The cross-normalizations added to the right panel of Fig. 2 also present in the dissipation range (with $K_d = \sqrt{k_{dn}k_{dh}}$ for normalization)

$$R_h/E_n \propto k^{-12/5} \text{ and } R_n/E_h \propto k^{-14/15}. \tag{20}$$

This observation means that,

$$F_h^R(k) = F_h^E(k),\ F_n^R(k) = F_n^E(k) \tag{21}$$

$$\text{and } \alpha_h^R - \alpha_n^E = -12/5,\ \alpha_n^R - \alpha_h^E = -14/15. \tag{22}$$

Thus, we deduce

$$F_h^R(k) = F_h^E(k) = F_n^R(k) = F_n^E(k) \tag{23}$$

$$\text{and } \Delta\alpha = \alpha_n^E - \alpha_h^E = 11/15 = \alpha_n^R - \alpha_h^R, \tag{24}$$

with all exponents determined up to a constant $\gamma$.

[The precise figures of the "measured" exponents in the above are obtained in an error-and-trial way with some iterations to reach the overall best fits, as shown by the respective straight lines in the figures, to all the data.] We remark that since $E_|$ in this range is much smaller than $R$, thus $\alpha_n^E - \alpha_h^E$ we deduced in the above is essentially for the transversal modes of the velocity fluctuations (more on the scalings of the helically decomposed spectra below).

Evaluation of the common part, $k^\gamma F(k)$, by fitting the spectra however can be subtle due to unknown *ansatz* for $F(k)$,[23] and is not of our current interest.

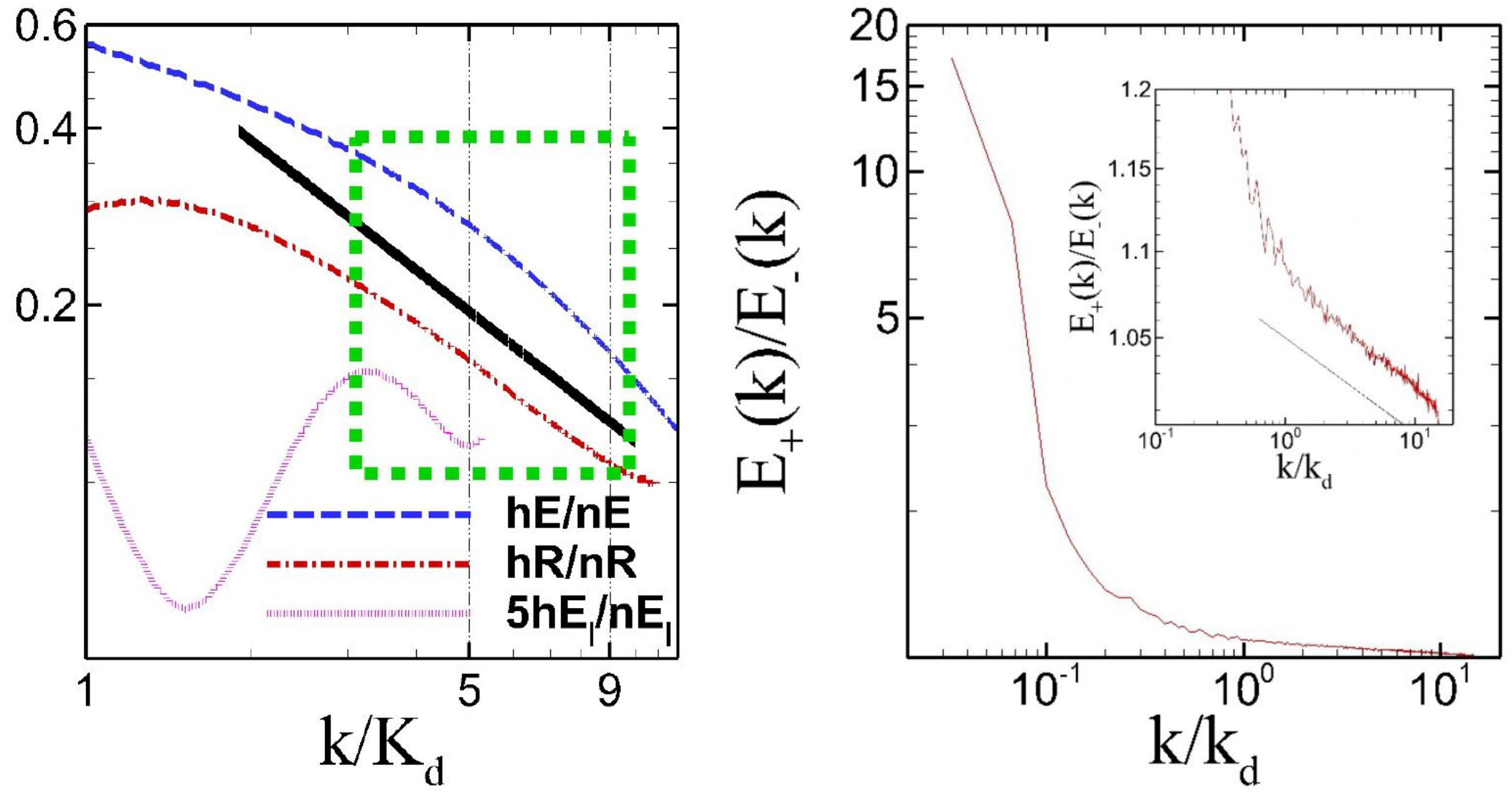

FIG. 3. Left panel: $E_h(k)/E_n(k)$ ('hE/nE') and $R_h(k)/R_n(k)$ ('hR/nR') and $5E_{|h}(k)/E_{|n}(k)$ ('5hE$_|$/nE$_|$'); Right panel: $E_+(k)/E_-(k)$ for the helical case.

Eq. (24) can also be verified, *a posteriori*, by $E_h(k)/E_n(k)$ and $R_h(k)/R_n(k)$ and even extended to characterize $E_|$, with however much less obvious scaling behaviors to be used for derivation from the beginning, due to the errors in our data: Fig. 3 plots in the left panel $E_h(k)/E_n(k)$ ('hE/nE') and $R_h(k)/R_n(k)$ ('hR/nR'), with the short straight line denoting the power law $\propto k^{-\Delta\alpha}$, verifying *a posteriori* the result obtained from Fig. 2 and indicating the extension to the parallel-mode spectra $(5\times)E_{|h}(k)/E_{|n}(k)$ ('5hE$_|$/nE$_|$') in the short range inside the dashed rectangular (the parallel-mode spectra, being much smaller than the density spectra, have at large $k$ less reliable data not seriously contaminated by numerical errors); also plotted in the right panel is the ratio of the helically decomposed spectra, presenting accurately (as designated by the straight short line)

$$E_+(k)/E_-(k) \propto k^{-\Delta\beta} \tag{25}$$

with $\Delta\beta = 1/45$ in the 'tail' (inset) for the helical case and indicating that the prefactor of $E(k)$ in our data actually is contributed by two slightly different powers of $k$ of comparable magnitudes (thus no clean power-law): the asymptotic subdominant prefactor actually has an exponent very close to the dominant one, so our previous results are accurate enough, with $\Delta\beta \ll \Delta\alpha$, for our purpose.

[With $\Delta\beta \ll \Delta\alpha$, $\Delta\alpha$ measured earlier is then accurate enough for our purpose. Actually, so small a $\Delta\beta$ cannot be very reliable as a linear fit (in the log-log plot), in the sense that, given the same absolute error, the relative error is larger with smaller $\Delta\beta$. And, Eq. (25) in terms of power law with a positive $\Delta\beta$ cannot be precise for $k \to \infty$, since in principle $E_+/E_- \searrow 1$, i.e., bounded from below by 1, with positive helicity injected at large scales. Thus, Eq. (25) is understood as an approximation (or some asymptotics of a slowly decaying function in an intermediate interval), which however does not exclude the possibility of being precise for other results associated to $\Delta\alpha$.]

A positive $\Delta\alpha$ indicates "effectively" (assembling the nature, distances to the real world and possible interactions) stronger complex singularities,[22] of the discretized system or asymptotically the NSE, for the nonhelical case. It is possible to make the analysis more definite with additional assumptions of the nature and distribution of the singularities[25] or more systematic techniques of singularity detection (highly subtle in high dimensions[26]), which however is beyond the scope of this note.

## V. EXPECTATION

There are two opposite but fundamentally connected directions for future investigations, *i.e.*, exposing even more basic dynamical mechanisms leading to the above statistical results and, as already remarked in the introductory discussions, applying the results in relevant practices such as those involving flow noise in civil engineering, rapid public transportation and car transportation, and aeronautics including the aircraft design. To motivate further discussions, we also present in Fig. 4 the typical snapshots of the normalized density-mode power spectra to show the fluctuations around the means in Fig. 2. The results, representative for others at different instants (not shown), clearly show fluctuations of the 'benefit' from helicity injection, much larger at the scales around those forced than at small scales (across the inertial and dissipation scales). Similar behavior presents in the snapshots of $E_|$ (not shown). The fact that more stable benefits are in the smaller scales, into which both (kinetic) energy and helicity transfer (from the forced large scales) and where helicity is much smaller, is probably an effect of 'self-averaging' (more effective for smaller scales with more spatial samples) at each instant.

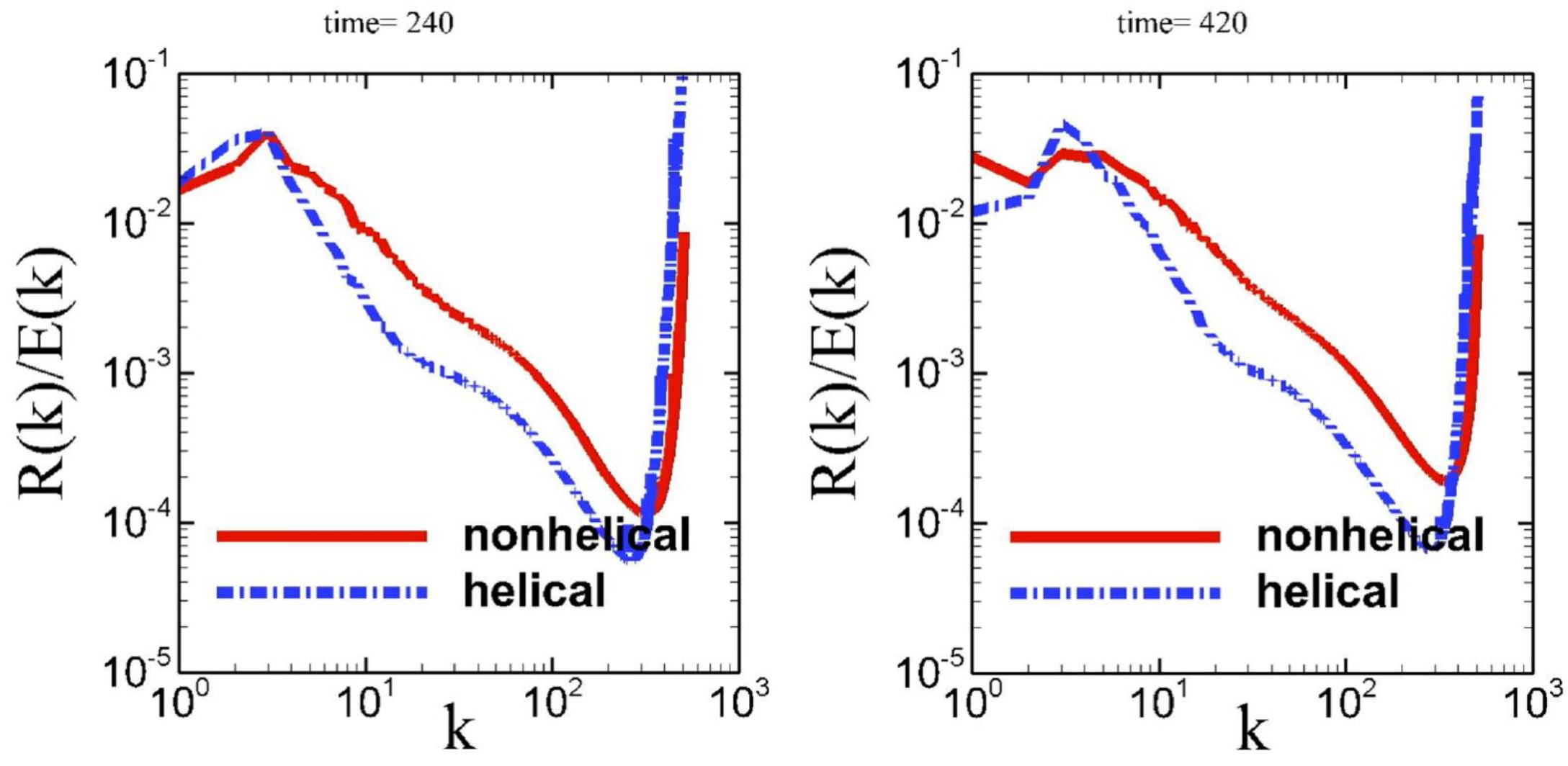

FIG. 4. Snapshots at two different times (240 and 420), among many others (not shown), of the original raw data (without wavenumber normalization) of $R(k)/E(k)$ showing fluctuations around the means in Fig. 2. The "hooks" at the large $k$ ends are due to the numerical errors and should be ignored (removed in the previous plots).

Thus, in the 'downward' basic direction, it would be even more illuminating to look into more quantitatively at different scales such effects of different values of helicity or degrees of reflexional symmetry breaking of the flow. A survey of different forcing scales can be helpful. Also, the analyses in the dissipation range of Sec. IV are for the data from the discrete dynamical system defined by the specific numerical scheme applied, and other numerical schemes (such as those applied in Ref. 8, among others) and methods are expected to be used for checking the possible universality of any sort, including the asymptotic behavior of the NSE: in principle, the grid spacing of our simulations should be small enough to resolve a reasonable span of $k$ in the dissipation range, but we indeed do not have a systematic nonlinear numerical analysis result to estimate the possible influence on the measured scaling exponents by the back scattering of the unresolved-$k$ modes (for the aliased $k$ up to 512, particularly left in Fig. 4), the latter depending on the algorithm and discretization scheme. Reflecting upon the state-of-the-art of the results, with no complete success of determining the precise shape of the spectra, of incompressible turbulence with spectral accuracy in the literatures of Ref. 23, we expect that three-dimensional Burgers equation should be able to help (preliminary results in Ref. 17 with much lower Reynolds number comfirm the current test of compressibility reduction with helicity by even a much larger amount of savings.)

While, in the 'upward' application direction, in the aircraft design optimization with the objective of noise reduction, say, at least two useful indications can be derived from the results. First, to have as more and stable benefits as possible, we should try to control the flow with specific design at scales as large as possible, and that the optimization through a single local small region at the rear of the wing may not be as efficient as that through multiple locations at the cutting edges to result in more spatially global effects. Second, Fig. 4 shows that the fluctuations of benefic at large scales (small $k$) can even be strongly negative, which indicates that the helicity control involves the potential systems penalties and we should try to make such fluctuations as small as possible; otherwise, the negative benefit could lead to even higher peaks/spikes of noise, though the mean appears still smaller with helicity. The latter may be relevant in taking both sound pressure level and annoyance into account for considering the various noise sources in engine system design.[27] Of course, a lot of practical studies, along with other fundamental supporting ones including the establishment of solid relations (e.g., Ref. 28), are still left for future, because, though there are many ways to perform flow control, far fewer have net system savings.

## ACKNOWLEDGMENTS

This work was partially supported by NSFC (Nos. 11672102, 11602277). JZZ owes a debt of gratitude to Uriel Frisch for patiently introducing the complex singularities years ago, which somehow planted the seeds of the motivation for the discussions at the end of Sec. IV which may be extended with the continuation of the intermittency studies in J.-Z. Zhu [Chinese Phys. Letters **23**, 2139 (2006)] for further information. The authors thank Jin-Xiu Xu, for the kind facilitation and help in the application of the supercomputer resources offered by the National Supercomputing Wuxi Center, and Jun Peng for the help with the codes in the early stage of this work. Two of the anonymous referees have greatly helped improving the presentation of the work with critical comments.

## DATA AVAILABILITY

The data that support the findings of this study are available from the corresponding author upon reasonable request.

## REFERENCES

[1] R. Betchov. Semi-Isotropic Turbulence and Helicoidal Flows. Physics of Fluids, 4: 925, 1961.

[2] H. K. Moffatt, Magnetic Field Generation in Electrically Conducting Fluids. (Cambridge University Press, 1978).

[3] V. I. Arnold & B. A. Khesin, Topological methods in hydrodynamics. (Springer, 1998).

[4] J.-Z. Zhu, Isotropic polarization of compressible flows. J. Fluid Mech., **787**, 440 (2016).

[5] R. H. Kraichnan, On the statistical mechanics of an adiabatically compressible fluid. J. Acoust. Soc. Am. **27**, 438 (1955).

[6] J.-Z. Zhu, Compressible helical turbulence: fastened-structure geometry and statistics. Phys. Plasmas, 28, 032302 (2021).

[7] J.-Z. Zhu, "The geometry and physics of the Taylor-Proudman theorems in $\mathbb{E}^{d\ (\geq 3)}$ and the plasma-flow analogues." arXiv:1905.11783 [math.AP]

[8] J.-Z. Zhu, Real Schur flow computations, helicity fastening effects and Bagua-pattern cyclones, *Phys. Fluids*, **33**, 107112 (2021).

[9] In the nonlinear, especially turbulent, regime, the notion of 'aeroacoustics' and 'noise' becomes subtle [c.f., the spatio-temporal spectra in, e.g., recent work of J. Cerretani and P. Dmitruk, Coexistence of acoustic waves and turbulence in low Mach number compressible flows. Phys. Fluids **31**, 045102 (2019)]. This note will not make specific identification of 'turbulence noise', but only refer to the latter by associating it to the compressibility of the flow.

[10] W. Devenport, N. Alexander, S. Glegg, and M. Wang, The Sound of Flow Over Rigid Walls. Annu. Rev. Fluid Mech. **50**, 435 (2018).

[11] J. W. Jaworski and N. Peake, Aeroacoustics of Silent Owl Flight. Annu. Rev. Fluid Mech. **52**, 395 (2020).

[12] Jameson, A. Computational Aerodynamics for Aircraft Design. Science, 245, 361–371 (1989).

[13] P. Yu, J. Peng, J. Bai, X. Han and X. Song, Aeroacoustic and aerodynamic optimization of propeller blades. Chinese Journal of Aeronautics, 33, 826–839 (2020).

[14] J. Shebalin, "Pseudospectral simulation of compressible turbulence using logarithmic variables," 11th Computational Fluid Dynamics Conference, Orlando, FL, U.S.A (1993).

[15] Collaboration et al., The pencil code, a modular MPI code for partial differential equations and particles: Multipurpose and multiuser-maintained, J. Open Source Software 6, 2807 (2021).

[16] J.-Z. Zhu, "Thermodynamic and vortic structures of real Schur flows," J. Math. Phys. **62**, 083101 (2021).

[17] J.-Z. Zhu Fastened compressible helical turbulence and component-wise dimensionally reduced flows, unpublished (2022).

[18] R.-H. Kraichnan, Helical turbulence and absolute equilibrium. J. Fluid Mech. **59**, 745 (1973).

[19] F. Waleffe, The nature of triad interactions in homogeneous turbulence. Phys. Fluids A **4**, 350–363 (1992).

[20] A. Brissaud, U. Frisch, J. Leorat, M. Lesieur, and A. Mazure, Helicity cascades in fully developed isotropic turbulence, Phys. Fluids **16**, 1366 (1973).

[21] A. Brandenburg, The Inverse Cascade and Nonlinear Alpha-Effect in Simulations of Isotropic Helical Hydromagnetic Turbulence. Astrophys. J. **550**, 824 (2001).

[22] This is a spectrum initially proposed for incompressible turbulence phenomenology, an indepth modern analysis of which can be found in U. Frisch, Turbulence: The Legacy of A. N. Kolmogorov. (Cambridge University Press, 1995).

[23] Relevant information of fluids, plasmas and beyond can be found in recent publications by, e.g., D. Buaria and K. R. Sreenivasan, Dissipation range of the energy spectrum in high Reynolds number turbulence, Physical Review Fluids **5**, 092601(R) (2020); A. Gorbunova, G. Balarac, M. Bourgoin, L. Canet, N. Mordant, and V. Rossetto, Analysis of the dissipative range of the energy spectrum in grid turbulence and in direct numerical simulations, Phys. Rev. Fluids 5, 044604,2020; S. Khurshid, D. Donzis and K. R. Sreenivasan, Energy spectrum in the dissipation range, Physical Review Fluids **3**, 082601 (2018); J. E. Maggs and G. J. Morales, Generality of Deterministic Chaos, Exponential Spectra, and

Lorentzian Pulses in Magnetically Confined Plasmas, Phys. Rev. Lett. 107, 185003 (2011); and their bibliographies.

[24]B. Dubrulle, Beyond Kolmogorov cascades. J. Fluid Mech. **867**, P1 (2019).

[25]U. Frisch and R. Morf, Intermittency in nonlinear dynamics and singularities at complex times. Phys. Rev. A, 23, 2673 (1981).

[26]F. Gargano, M. Sammartino, V. Sciacca, and K. W. Cassel, Analysis of complex singularities in high-Reynolds-number Navier-Stokes solutions, J. Fluid Mech. **747** 381 (2014); W. Pauls and U. Frisch, A Borel transform method for locating singularities of Taylor and Fourier series, J. Stat. Phys. **127**, 1095 (2007); R.E. Caflisch, F. Gargano, M. Sammartino, V. Sciacca, Complex singularities and PDEs, Riv. Mat. Univ. Parma **6**, 1 (2015).

[27]X.-F. Xin, Noise, vibration, and harshness (NVH) in diesel engine system design, in *Diesel Engine System Design*; Editor(s): Qianfan Xin; Woodhead Publishing, 759–-821 (2013).

[28]S. Kurien, The reflection-antisymmetric counterpart of the Kármán–Howarth dynamical equation, Phys. D **175**, 167–-176 (2003).